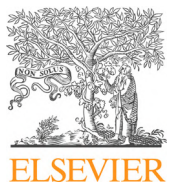

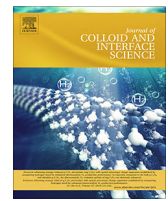

Regular Article

# Impact of surface curvature, grafting density and solvent type on the PEGylation of titanium dioxide nanoparticles

Daniele Selli [a], Stefano Motta [b], Cristiana Di Valentin [a],*

[a] Dipartimento di Scienza dei Materiali, Università di Milano-Bicocca, via R. Cozzi 55, I-20125 Milano, Italy
[b] Dipartimento di Scienze dell'Ambiente e della Terra, Università di Milano-Bicocca, Piazza della Scienza 1, I-20126 Milano, Italy

G R A P H I C A L   A B S T R A C T

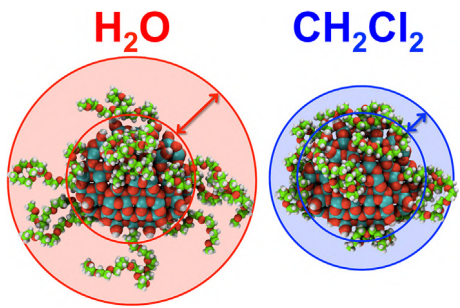

A R T I C L E   I N F O

A B S T R A C T

TiO$_2$ nanoparticles (NPs) are attracting materials for biomedical applications, provided that they are coated with polymers to improve solubility, dispersion and biocompatibility. Conformation, coverage density and solvent effects largely influence their functionality and stability. In this work, we use atomistic molecular dynamics simulations to study polyethylene glycol (PEG) grafting to highly curved TiO$_2$ NPs (2–3 nm) in different solvents. We compare the coating polymer conformations on NPs with those on (1 0 1) flat surfaces. In water, the transition from mushroom to brush conformation starts only at high density ($\sigma = 2.25$ chains/nm$^2$). In dichloromethane (DCM), at low-medium coverage ($\sigma < 1.35$ chains/nm$^2$), several interactions between the PEG chains backbone and undercoordinated Ti atoms are established, whereas at $\sigma = 2.25$ chains/nm$^2$ the conformation clearly becomes brush-like. Finally, we demonstrate that these spherical brushes, when immersed in water, but not in DCM, follow the Daoud-Cotton (DC) classical scaling model for the polymer volume fraction dependence with the distance from the center of star-shaped systems.

© 2019 Elsevier Inc. All rights reserved.

## 1. Introduction

Nanosized TiO$_2$ particles are commonly used in different technological fields, ranging from dye-sensitized solar cells (DSSC) [1,2] to photocatalysis and to photoelectrochemistry [3–7]. Following excessive dilution, TiO$_2$ nanocrystals edges dissolve and spherical nanoparticles are formed [8]. Due to their highly curved surface, which exposes undercoordinated atoms [8–10], these nano-objects are extremely reactive and can be easily functionalized [11–14]. For this reason, spherical TiO$_2$ nanoparticles have gained major attention for applications in nanomedicine; in particular, they are perfect building block of novel bioinorganic hybrid nanoconjugates, which show enhanced features for imaging, photocatalytic therapy and photoactivated drug release [15–20].

* Corresponding author.
  E-mail addresses: daniele.selli@unimib.it (D. Selli), stefano.motta@unimib.it (S. Motta), cristiana.divalentin@unimib.it (C. Di Valentin).

However, it is well known that bare spherical $TiO_2$ nanoparticles (NPs) are cytotoxic and that they have to be pre-treated for clinical usage [17,21–23]. The reduction of toxicity can be achieved by decorating the NP surface with polymer chains, which affects other NP properties, such as solubility, mobility, size and tissue penetration [24–26]. Among a wide range of biocompatible and biodegradable polymers used to cover and protect nanoparticles surface from the surrounding environment, polyethelene glycol (PEG) is probably the most used one due to a long list of advantages [27]. PEG is approved by the U.S. Food and Drug Administration (FDA), inexpensive, improves hydrophilicity, prevents agglomeration and opsonization of the nanoparticles allowing for a favorable pharmacokinetics and tissue distribution [28–31]. Furthermore, if a sufficiently high grafting density is achieved, the uptake by the reticuloendothelial system (RES) is avoided and the in vivo circulation time increases [32,33].

It appears clear that a high grafting density is desirable to achieve complete coverage and improve biocompatibility. For this reason, different grafting routes have been explored during the last years, especially regarding metallic (gold) and metal oxide ($TiO_2$, $SiO_2$ and $Fe_3O_4$) nanoparticles. The "grafting to" is usually preferred to the "grafting from" procedure, since it has been shown to provide better coverage, enhanced miscibility and lower propensity to aggregate of the grafted nanoparticles [34]. In any grafting process, two are the key factors: (i) the choice of the linker used to chemically attach the polymer to the surface and (ii) the choice of the solvent in which the grafting process takes place during the synthesis. Thiol/silane [35,36], catechol [37,38], carbonyl/carboxyl [15] or phosphate [39] modified PEG are typically chosen as linker functionalities. Different solvents can be used as medium for the polymer grafting process on metal oxide nanoparticles, however water is the most popular one, due to the foreseen biomedical applications. Ma et al. [36] PEGylated silica nanoparticles in water; Yang et al. [40], Liu et al. [41] and Cano et al. [42] used the same solvent for PEGylation of iron oxide nanoparticles. Water is the preferred solvent also for $TiO_2$ nanoparticles functionalization, as reported in many different works [15,16,37,43].

However, several other studies show that functionalization of $TiO_2$ nanoparticles with organic molecules can be easily achieved in various other solvents. Gerrero et al. [14] successfully attached organophosphorus coupling molecules to titania NPs in methanol, toluene and dichloromethane (DCM). Weng et al. [44] investigated binding modes of the carboxylic terminated ATRA molecule to $TiO_2$ nanoparticles in methanol. In other studies, 30 nm wide dye-modified $TiO_2$ nanoparticles were prepared in DCM [45], while isopropanol has been recently used to obtain self-assembled monolayers (SAM), on $TiO_2$ NPs of about 34 nm of diameter, of several organic molecules containing phosponic, carboxylic and catechol ligand groups [46].

From a computational point of view, many studies have been published in the last years regarding chemically modified low index flat surfaces of $TiO_2$ [47–53], however, only few of them report on the modelling of functionalized nanoparticles [54–57]. In particular, simulation of polymer functionalization of NPs is a complex task and requires sophisticated tools. In the literature, spherical brushes have been studied with Monte-Carlo and coarse-grained molecular dynamics (MD) simulations [58–60]. PEGylation of gold NPs has been recently investigated both with coarse-grained and all-atoms MD [61,62], whereas classical MD runs have been carried out on PEG-coated silica NPs and on the interaction of iron oxides nanoparticles with different surfactants [63,64]. To the best of our knowledge, no studies on polymer-coated curved $TiO_2$ nanoparticles exist. Moreover, only in two of the works mentioned above, the effect of the solvent on the polymer chains conformation has been taken into account, either by using an implicit solvent treatment [60] or by means of coarse-grained techniques [61]. Therefore, an atomistic description on the role played by the solvent on the conformation of the grafted polymers is still lacking.

In this work, we use all-atom molecular dynamics to study how a free PEG chain and PEG-coated spherical anatase $TiO_2$ nanoparticles interact with the different surrounding media, namely water and dichloromethane (DCM). We focus on the conformational changes that the PEG chains in the $TiO_2$ spherical brushes (PEG@NPs) undergo with increasing coverage densities and in different solvents. Peculiar differences are observed in the spatial distribution of the polymer around the NPs in water or in DCM. This analysis provides a rational basis for an improved design of PEG-grafted $TiO_2$ nanoparticles for biomedical applications.

The paper is organized as follow: in Section 2 we present the computational details on the methods and models used for the simulations (Section 2.1) and the indicators for the conformational analysis of the PEG-grafted $TiO_2$ NPs (Section 2.2). In Section 3 we present our results, first for the bare NP immersed in water and in DCM (Section 3.1), secondly for the free PEG in the same solvents (Section 3.2), then for a PEG-grafted anatase (1 0 1) $TiO_2$ flat surface (Section 3.3) and, finally, for the PEG@NP systems (Section 3.4). In Section 4, the results are summarized and relevant conclusions are given on the role of the solvent in designing next generation nanohybrids for medical applications.

## 2. Computational details

### 2.1. Models and methods

All the classical molecular dynamics (MD) calculations were carried out with the AMBER16 simulation package [65]. The simulations were performed in orthorhombic boxes with periodic boundary conditions imposed in all directions. The integration time step was set to 0.5 fs to ensure reversibility and the cutoff for the electrostatic interactions set to 1.2 nm. All the systems were subject to an initial minimization of 50.000 steps to avoid atoms overlap. Successively, the systems have been gradually heated up to the target temperature of 300 K in 500 ps using a Langevin thermostat ($\gamma = 1$ ps$^{-1}$) [66]. The equilibration run has been performed for 5 or 35 ns depending on the system, while the production run has been performed for 25 ns in the NVT ensemble for all the systems.

For water the TIP3P [67] model was used, while we used the Generalized Amber Force-Field (GAFF) [68] for dichloromethane (DCM) as reported in Ref. [69]. For $TiO_2$ we used a simplified version of the Matsui-Akaogi force-field [70] as reported by Luan et al. [71], while to describe the polyethylene PEG we used a standard procedure for the atomic charge calculation: (i) we optimized the monomer geometry with Gaussian16 [72] at the HF/6-31G* level of theory and obtained the charges according to the Merz–Kollman population analysis scheme [73,74]; (ii) we derived the partial atomic charges according to the RESP method [75] utilizing the ANTECHAMBER module [76]. Finally the topology of the whole system was generated with the tLEap module of AMBER16, according to GAFF. To simulate the PEG chain, we considered a methoxy-PEG, $H_3C-[OCH_2CH_2]_n-OH$, with n = 11 for a total molecular weight ($M_w$) of 516 au.

The spherical anatase $TiO_2$ nanoparticle model used throughout this work has been designed through global optimization with a simulated annealing process at the density functional tight binding (DFTB) [77] level of theory and, successively, optimized with the B3LYP density functional theory (DFT) method in a previous work by some of us [9]. The stoichiometry of the model is $(TiO_2)_{223} \cdot 10H_2O$ and it is characterized by an equivalent diameter of 2.2 nm. The ten water molecules are used to saturate superficial Ti or O atoms, which are too undercoordinated (2/3-fold or 1-fold, respectively). The details of the procedure used to prepare the NP models starting from the anatase bulk have been reported in a previous publication by some of us [10]. During the molecular dynamics simulations performed in the present study, the structure of the NP has been kept fixed to the DFT (B3LYP) optimized one, using very high Cartesian restraints (5000 kcal mol$^{-1}$ Å$^{-2}$). It is important to underline that the pH of the solution is a determining factor for the surface charge of $TiO_2$ nanoparticles [78,79]. In particular, at pH = 7 the NPs surface is found to be negatively charged, probably due to the presence of some superficial -$TiO^-$ groups [80], even if neutral charge is expected around pH = 6 [81]. The modeling of $TiO_2$ in different pH conditions is certainly extremely important, but it goes beyond the scope of this study. For this reason our NP model is neutral and the hydroxyl group on the surface of the nanoparticle are undissociated.

To analyze the NP/SOLVENT and PEG/SOLVENT interaction the NP and the PEG models have been immersed in a $6 \times 6 \times 6$ nm box of solvent (water or DCM) using the PACKMOL code [82]. The number of solvent molecules has been chosen in order to have densities of about 1.00 g/cm$^3$ for water and 1.24 g/cm$^3$ for DCM.

To functionalize the spherical $TiO_2$ nanoparticle, linear PEG chains have been bonded first to the 4-fold coordinated Ti atoms of the surface, which we have recently reported [9] to be the most reactive ones, and then to the 5-fold coordinated Ti atoms, which allowed for a certain symmetry of the system. We considered seven different coverage densities for the PEG@NP system in water and in DCM, $\sigma$ = 0.225, 0.440, 0.676, 0.901, 1.126, 1.351, 2.252 chain/nm$^2$ (all the models are reported in Fig. S1 in the Supporting Material). These densities and the relative number of PEG chains attached to the surface are reported in Table 1. DFT calculations on the adsorption of PEG dimers on the anatase (1 0 1) surface of $TiO_2$ [51] proved that the adsorption process is exothermic, with binding energies from about $-1.0$ to $-2.0$ eV depending on the adsorption mode and on the grafting density. Therefore, after functionalizing the NP, we performed an energy minimization and then we fixed the distance between the undercoordinated Ti atoms (involved in the binding with the PEG) and the terminal O atoms of each PEG. Successively, the PEG@NP system has been immersed in a $10 \times 10 \times 10$ nm box of solvent (water or DCM).

For comparison we have simulated also the PEG grafting on the anatase $TiO_2$ (1 0 1) flat surface. The surface was modeled with a $4 \times 6$ supercell of three triatomic layers of $TiO_2$. The model has a total of 864 atoms and 48 undercoordinated Ti$_{5c}$ (5-fold coordinated) adsorption sites. The so-called *mushroom* reference structure has been obtained by grafting a single PEG chain, while the *brush* reference one by adsorbing 24 PEG chains, for a resulting grafting density of $\sigma$ = 0.107 and 2.573 chain/nm$^2$, respectively (see Table 1). The volume between periodic replicas of 52 Å height along the z-axis has been filled with solvent molecules for densities of about 1.00 g/cm$^3$ for water and 1.24 g/cm$^3$ for DCM.

**Table 1**
Number of PEG chains attached to the anatase $TiO_2$ (1 0 1) surface and to the NP and corresponding grafting densities, $\sigma$ (chain/nm$^2$).

| System | N. of chains | $\sigma$ (chain/nm$^2$) |
| --- | --- | --- |
| Mushroom | 1 | 0.107 |
| PEG@NP | 5 | 0.225 |
| | 10 | 0.440 |
| | 15 | 0.676 |
| | 20 | 0.901 |
| | 25 | 1.126 |
| | 30 | 1.351 |
| | 50 | 2.252 |
| brush | 24 | 2.573 |

## 2.2. Indicators for the analysis

Several quantities have been evaluated to describe the NP/PEG/SOLVENT interactions and the PEG chains conformation. For all the PEG@NP and *brush* systems each value reported below has been averaged between 2000 configurations collected every 2.5 ps during the last 5 ns of the production run. In the case of the free PEG and the *mushroom* systems, we have performed three different production runs to increase the statistics.

The indicators used for the analysis are the following:

(i) *Distribution of distances:* g(d) is the distribution function of water or DCM around the spherical nanoparticle. *d* is defined as the distance of the water O atom (or the DCM C atom) from the closest surface Ti atom.

(ii) *End-to-end distance:* $\langle h^2 \rangle^{1/2}$ is the distance between the first and the last heavy atom (oxygen of the —OH head and carbon of the —CH$_3$ tail) of the PEG chain.

(iii) *Radius of gyration:* R$_g$ is the root-mean-square distance between each atom in the molecule and the center of mass of the molecule itself, normalized for the number of atoms.

(iv) *OCCO dihedral angles distribution:* it is the distribution of the value of all the OCCO dihedral angles for all the PEG chains considered. We define *trans* the angles with values <$-150°$ or >$150°$, *anticlinal* the ones with values between $\pm150°$ and $\pm90°$, *cis* the ones with values >$-30°$ and <$30°$ and gauche (+)/gauche (−) the ones with values between $\pm90°$ and $\pm30°$. Each PEG chain considered has 10 dihedral angles.

(v) *H-bonds:* average number, per monomer in the chain, of hydrogen bonds between water and the O atoms of the PEG chains. We consider the formation of a H bond when the geometrical criteria of distance r$_{OO}$ < 3.5 Å and angle DHA (donor–hydrogen-acceptor) <150° are encountered.

(vi) *Mean distance from the surface (MDFS):* it is the average distance of the center of mass of each PEG chain grafted to the NP and the closest Ti atom of the surface.

(vii) *O$_{PEG}$-Ti bonds:* number of bonds between the O atoms of the PEG chains and the undercoordinated Ti atoms of the NP surface, normalized for the number of grafted PEG chains. Even though molecular mechanics does not allow for the formation of real bonds, we consider the formation of a bond when distance between the O atoms of the PEG chain and the Ti atoms is lower than 2.5 Å due to an electrostatic interaction.

(viii) *OCCO dihedral index (DAI):* this index is calculated as follows. The angles between $\pm90°$ (corresponding to gauche (+)/gauche (−)) and *cis* dihedrals) of the OCCO dihedral distributions are removed in order to highlight the differences only in the *anticlinal* and *trans* distribution range. Then, everything is integrated between $-180°$ and $+180°$. Therefore, large values of DAI indicates a high number of *anticlinal* and *trans* dihedral angles.

(ix) *PEG and Solvent volume fractions:* the volume fractions, $\Phi$(r), of PEG and solvent are calculated for the last 100 snapshots of the MD simulations using spherical layers of 0.09 nm starting from the center of the NP. For each CH$_2$ unit and O atom of the PEG we used a volume of 0.02 nm$^3$, for water of 0.03 nm$^3$ and for DCM of 0.06 nm$^3$.

To assess the equilibration of our simulations and the convergence of the MD indicators, we analyzed the values of the *radius of gyration*, *end-to-end distance* and number of *H-bonds* per monomer along the equilibration (5 or 35 ns) and production (25 ns) runs, as shown for some selected case studies in Fig. S2 (i.e. PEG@NP systems with $\sigma$ = 0.225 and $\sigma$ = 2.252 in water and DCM). We plotted the average value over every 5 ns fragments of the MD trajectory as red squares in the graphs together with the standard deviation range within the red bars. Since the mean value of the indicators along the entire production run, represented by a green line, is contained within the red bars of all the entries, we conclude that the indicators reached a proper convergence.

## 3. Results and discussion

### 3.1. A bare TiO₂ nanoparticle in water and in dichloromethane

Starting our analysis from the aqueous environment, a crucial aspect to investigate is the effect of the highly curved surface of this nano-object on the water structure. In Fig. 1a we report the distribution function $g(d)$ of the water around the TiO₂ NP (position of the NP atoms are kept fixed at the optimized DFT(B3LYP) geometry) [9]. We register three main peaks: the first at around 2.25 Å refers to water molecules forming coordinative bonds with the undercoordinated Ti atoms on the NP surface, as it can be also inferred from the narrow broadening of the peak. This is in line with both quantum methods (DFT) and experiments, showing that on curved NP water molecules molecularly adsorb or even partially dissociate [57,83]. The second peak at ~3.8 Å refers to water molecules that are H-bonded to O atoms of the NP surface, while the third peak at 4.45 Å refers to those that are H-bonded to Ti-OH or Ti-H₂O through the O$_{water}$ atoms. Beyond this distance water rapidly behaves similarly to bulk (already at 6.0 Å the $g(d)$ is about 1).

The long-range effect of the TiO₂ nanoparticles surface on water has been already shown in other classical molecular dynamics study [54,56,84]. In particular, it has been reported in a recent publication by some of us [57], that the effect is propagated even to longer distances if a correct quantum chemical explicit description of the solvent, with a DFT-based method, is employed. Our simulation clearly shows that water tends to gather around the NP surface, increasing its density up to a distance of, at least, 6.0 Å.

In Fig. 1b we analyze how the TiO₂ NP curved surface influences the dynamical behavior of the dichloromethane, reporting the distribution function $g(d)$ of the solvent at increasing distance from the surface. The DCM density profile is very different compared to that of water (Fig. 1a). We observe a quite high peak at 4.15 Å corresponding to the first solvation shell. Then, between 7.2 Å and 9.5 Å the density presents another peak, lower than the previous one, and, finally, at around 12.5 Å there is one more smooth feature.

Thus, the presence of the curved NP in the solvent causes a much longer-range effect on the density of DCM (~12.5 Å) compared with what observed for water (~6.0 Å). Furthermore, the fact that the position of the first peak of the DCM distribution function $g(d)$ is at 4.15 Å, much further than for water (at 2.25 Å), is the direct consequence of the large van der Waals radii of the C and Cl atoms. In addition, the larger broadening of the same peak gives a clear indication that there is no interaction between the NP surface and the solvent molecules. This will have crucial consequences on the conformation of the PEG@NP system, as we will see in the following sections.

### 3.2. A free PEG chain in water and in DCM

The conformational analysis and the investigation of the dynamic behavior for a free PEG chain in a box of solvent molecules are essential, on one side, to validate the generated force-fields and, on the other, to build up a set of reference values for the indicators that will be used in the comparison with the PEG chains anchored to the TiO₂ NP.

Water is a H-bond donor solvent. One might expect that it should easily dissolve PEG. However, it is not the best solvent for PEG polymer chains. Others, such as methanol, THF and chloroform, are found to better solvate PEG [85]. It has been shown [81,86] that chloroform is an aprotic theta (Θ), thus weakly interacting, solvent with respect to PEG. We expect dichloromethane (DCM) to behave similarly. However, a complete study of PEG solubility in water and DCM is certainly beyond the scope of this work and, therefore, we will report only some indicators, which are important to understand the differences between the free polymer and the polymer grafted on the TiO₂ spherical nanoparticle.

Very important structural parameters for the polymer conformational analysis are the *end-to-end distance* ($\langle h^2 \rangle^{1/2}$) and the *radius of gyration* ($R_g$). In Table 2, we report the average root-mean-square value for these two quantities in water and we compare them with previous molecular dynamics simulations [87–89] and with experimental data obtained from static light scattering (SLS) measurements [90,91]. The agreement is rather satisfactory.

Furthermore, to highlight the dynamic structural changes of the PEG chain in water, we evaluate the conformation of the polymer considering the variation of each *OCCO dihedral angle*. The time averaged distribution of dihedral angles reported in Fig. 2a shows that the vast majority of the angles are characteristic of gauche (−) or gauche(+) conformations, i.e. between ±30.00° and

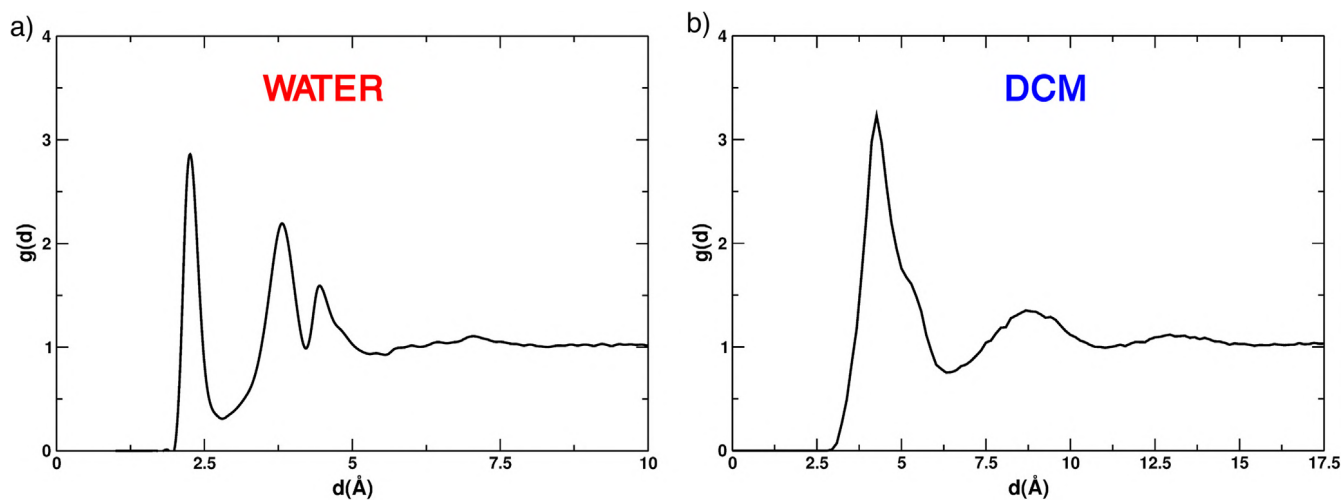

**Fig. 1.** Distribution function $g(d)$ as extracted from the molecular dynamics runs of a) water and b) DCM molecules surrounding the spherical TiO₂ NP model.

**Table 2**

Values of the *radius of gyration*, $R_g$ and *end-to-end distance*, $\langle h^2 \rangle^{1/2}$ (in nm) in water, as calculated in this work and from other computational and experimental studies. N refers to the number of monomers present in the PEG chain.

| Reference | N | $R_g$ (nm) | $\langle h^2 \rangle^{1/2}$ (nm) |
|---|---|---|---|
| This work | 11 | 0.60 ± 0.08 | 1.43 ± 0.40 |
| MD, Ref. [83] | 10 | 0.62 ± 0.09 | 1.43 ± 0.51 |
| MD, Ref. [84] | 9 | 0.63 ± 0.10 | 1.51 ± 0.40 |
| MD, Ref. [85] | 10 | 0.62 | 1.57 |
| [a]Exp, Ref. [86] | 11 | 0.81 ± 0.17 | – |
| [b]Exp, Ref. [87] | 11 | 0.62 | – |

[a] Calculated according to the relation $R_g = 0.0215 \, M_w^{(0.583 \pm 0.031)}$.

[b] Calculated according to the relation $R_g = 0.0202 \, M_w^{0.550}$.

±90.00°, even if there is a significant number of angles with higher values (till ±120.00°). The distributions are peaked at around −70.50° and +71.50° and the amount of (−) and (+) conformations is equally distributed. This is in good agreement both with the MD study of Oelmeier et al. [92], where they report a distribution equally peaked at −73.43° and +74.14° and with experimental crystallographic data [93], which indicates −74.95° as gauche dihedral angle for the PEG molecular crystal.

Finally, another important aspect to consider of the interaction between the PEG polymer chain and a polar solvent, such as water, is the total number of H-bonds that can be established between the solute and the solvent. The average number of *H-bonds* per monomer has been calculated to be 0.99 ± 0.08, which is slightly smaller than 1.08 obtained by Oelmeier et al. for a PEG chain of comparable length or 1.2 calculated by Dahal et al. for a 36 monomers PEG [88,94]. This indicator is particularly relevant since it is directly related to the size and conformation of the polymer. High molecular-weight PEG chains allow form a smaller number of H-bonds since oxygen atoms tend to face inwards and are excluded from interactions with the solvent. In addition, it has been demonstrated that stretching the PEG chain ruptures water bridges (water molecules H-bonded to two different oxygen atoms of the same PEG chain) and, consequently, reduces H-bonding with water [88,95].

The situation is extremely different when PEG is immersed in dichloromethane. The average root mean square *end-to-end distance* and *radius of gyration* correspond to 0.34 ± 0.03 nm and 0.51 ± 0.02 nm, respectively. The $\langle h^2 \rangle^{1/2}$ value is extremely small indicating that the polymer chain tends to coil and form an intramolecular H-bond between the hydrogen atom of the —OH head and one oxygen atom of the polymer chain, instead of interacting with the solvent (see Fig. S3 of the Supporting Material). This is confirmed from the relatively small value of the $R_g$, which is what one would expect for a weakly interacting solvent [96].

In Fig. 2b the *OCCO dihedral angle* distribution for a PEG chain in DCM is reported. The chain is mostly coiled during our simulations, whereas in water it is much more mobile and open. Although the vast majority of the angles are in gauche(−) or gauche(+) configuration, i.e. between ±30.00° and ±90.00°, we observe several broader angles (till ±120.00°), more than in the case of water (see Fig. 2a). The distributions are peaked at around −79.32° and +80.29° therefore about 10° larger than in water. Also in this case, the (−) and (+) conformations are almost equally distributed. We must note that, due to Θ nature of DCM, intramolecular interactions are promoted, leading to the formation of linear segments during the MD simulations (see Fig. S3 of the Supporting Material). This can be noticed comparing the values of the *OCCO dihedral angle* distribution close to *trans* (±180.00°): it is evident that in DCM (see inset of Fig. 2b) there is a higher number of these angle values compared to water (see inset of Fig. 2a). To give a quantitative evaluation of the conformation of the free PEG in the two solvents, we have calculated the *OCCO dihedral angle index* (DAI). As reported in details in Section 2.2, the larger is this index, the more abundant are the *OCCO dihedral angles* close to the *trans* value. The DAI for the free PEG in water is 0.13, while in DCM the same index is more than double, i.e. 0.28, in line with what we have just discussed about the conformation of the polymer in this Θ solvent.

### 3.3. PEG grafted to the (1 0 1) anatase TiO₂ surface in water and in DCM

As already highlighted in other studies [62], the role of the NP curvature is fundamental for determining the maximum possible grafting density of a polymer such as PEG. For this reason, we will compare our results for the PEG@NP systems not only with the free polymer in water and DCM, but also with two differently PEG-grafted flat anatase (1 0 1) TiO₂ surfaces that will be presented in this section (see Fig. 3). These two models represent the extreme cases of lowest and highest grafting densities and will be used as reference systems in the next section: (1) a single isolated PEG chain anchored on the surface (density of $\sigma$ = 0.107 chain/nm²), which we will define as the "*mushroom*" model, and (2) twenty-four PEG chains anchored on the surface (density of $\sigma$ = 2.573 chain/nm²), which we will define as the "*brush*" model.

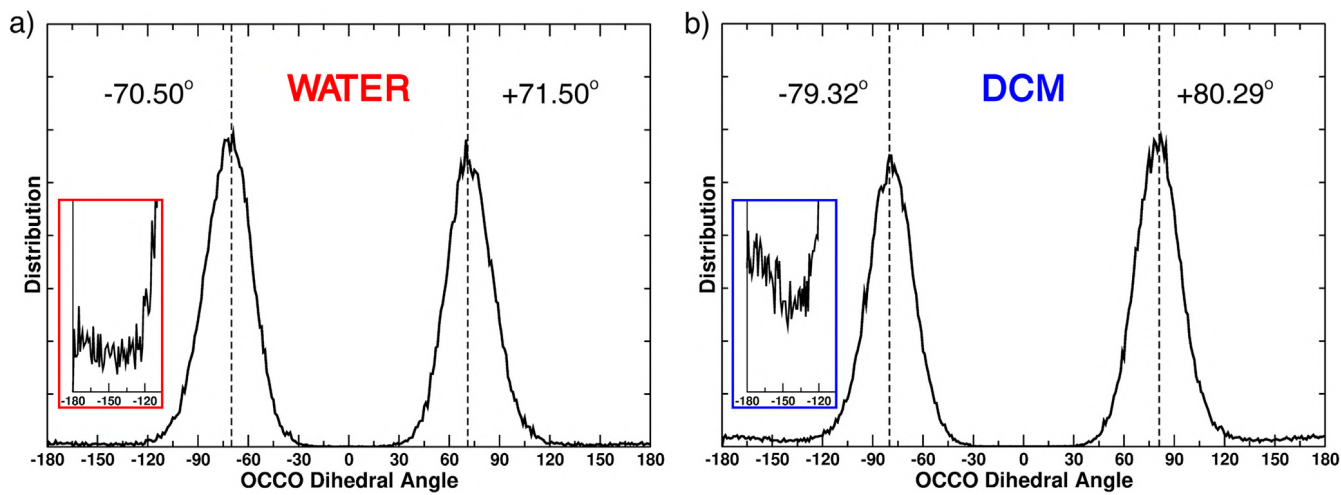

**Fig. 2.** (a) Averaged distribution of *OCCO dihedral angles* of the PEG chain in water. The majority of the angles have values between ±30.00° and ±90.00° corresponding to gauche(−/+) conformations. The two peaks are centered at −70.50° and +71.50°. (b) Averaged distribution of *OCCO dihedral angles* of the PEG chain in DCM. The majority of the angles have values between ±30.00° and ±90.00° corresponding to gauche(−/+) conformations. The two peaks are centered at −79.32° and +80.29°. For both plots, in the insets a magnification of the distribution close to the *trans* value is shown.

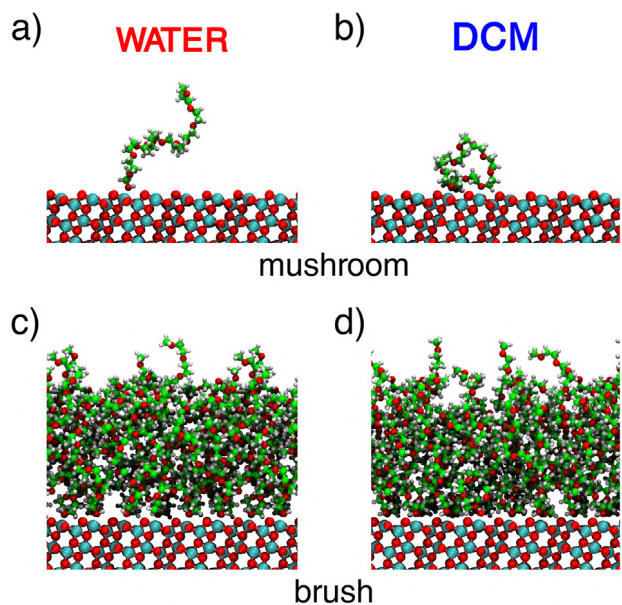

**Fig. 3.** Last MD snapshots of the *mushroom* and *brush* reference models of PEG grafted to the anatase (1 0 1) TiO₂ flat surfaces in water (a), (c) and DCM (b), (d). Titanium, oxygen, carbon and hydrogen atoms are in cyan, red, green and white, respectively. (For interpretation of the references to colour in this figure legend, the reader is referred to the web version of this article.)

**Table 4**
Values of the *mean distance from the surface* (MDFS) in nm of the *mushroom*, *brush* reference model and PEG@NP systems, immersed in water at different grafting densities ($\sigma$).

| N. of chains Grafted | Grafting density $\sigma$ (chain/nm²) | MDFS (nm) |
|---|---|---|
| Mushroom | 0.107 | 1.04 ± 0.18 |
| 5 | 0.225 | 1.00 ± 0.11 |
| 10 | 0.440 | 1.02 ± 0.06 |
| 15 | 0.676 | 0.91 ± 0.06 |
| 20 | 0.901 | 0.98 ± 0.03 |
| 25 | 1.126 | 1.04 ± 0.04 |
| 30 | 1.351 | 0.98 ± 0.03 |
| 50 | 2.252 | 1.11 ± 0.01 |
| Brush | 2.573 | 1.44 ± 0.01 |

For the *mushroom* and *brush* reference models, we have evaluated the *mean distance from the surface* (MDFS), the root mean square *end-to-end distance* $\langle h^2 \rangle^{1/2}$ and the *radius of gyration* $R_g$ as defined in Section 2.2 (see Table 3). In the case of water, we have also calculated the number of *H-bonds* per monomer in the chain, while for DCM, the number of $O_{PEG}$-Ti bonds per PEG chain (see Table 3). Values of these indicators are also reported in Tables 4–7 for comparison with free PEG and PEG@NP systems.

It is interesting to notice that at very low grafting density (*mushroom* model), the indicators measured in the two solvents are rather different. In DCM, the MDFS, $\langle h^2 \rangle^{1/2}$ and $R_g$ values are smaller than in water, suggesting that the polymer arranges closer to the surface in a more coiled conformation. This quantitative analysis is confirmed by the ball-and-stick structures of the two *mushroom* systems in water and DCM reported in Fig. 3a and b. Water establishes many *H-bonds* (1.01 ± 0.35 per monomer) with the chain and prevents its coiling, self-interaction and interaction with the surface. On the contrary, as we have already discussed in Sections 3.1 and 3.2, the PEG/DCM and TiO₂/DCM interactions are very weak, turning in the possibility for the polymer to reach the surface and form an average of 2.19 ± 0.68 $O_{PEG}$-Ti bonds per chain.

The situation is different at high grafting density (*brush* model). In this case, the values of the indicators, reported in Table 3 for water and DCM, are closer and the structures of the flat brushes look similar (see Fig. 3c and d). At this coverage the intramolecular PEG interactions become much more important than the PEG/SOL-VENT and PEG/TiO₂ ones, leading to a very low number of *H-bonds* per monomer (0.50 ± 0.04) in water and a value of $O_{PEG}$-Ti bonds per polymer (1.04 ± 0.06) close to 1 in DCM.

The evaluation of the *OCCO dihedral angle index* (DAI) gives us further information on the structure of the PEG chains grafted on the flat anatase (1 0 1) TiO₂ surface. Regarding the *mushroom* model in water, we have a relatively small DAI value of 0.14, very close to that of free PEG one in the same solvent (0.13). The *brush* model presents an index of 0.25 indicating an increase number of *angles* with values close to the *trans* one and, therefore, a stretching of the chains (typical of *brush-like* conformations). The *mushroom* and *brush* systems in DCM have surprisingly close DAI values, 0.29 vs 0.30, respectively. While in the *brush* case this is due to the expected stretching of the chains (as it happens in water), for the *mushroom* model the situation is different. As a matter of fact, despite the evident coiling of the polymer (see Fig. 3a), the formation of different $O_{PEG}$-Ti bonds forces the PEG chain to have a more abundant number of close-to-*trans* OCCO dihedrals than expected (see Fig. S4 in the Supporting Material).

It appears clear that, when PEG is grafted on a flat anatase (1 0 1) TiO₂ surface, at low coverage the solvent effect is still prominent and therefore two different *mushroom* conformations are expected. In water, the chain is completely immersed in the solvent and has a conformation similar to the free PEG (very close DAI values). In DCM, the weak PEG/SOLVENT and SOLVENT/TiO₂ interactions allow the chain to coil and bind the surface with different oxygen atoms of the backbone. At high coverage, when the solvent effect decreases consistently, even in water, intramolecular PEG interactions become prominent and the two *brush* models considered (in water and DCM) show similar behavior.

**Table 3**
Values of the *mean distance from the surface* (MDFS), *end-to-end distance* $\langle h^2 \rangle^{1/2}$ and *radius of gyration* $R_g$ (in nm) of the *mushroom* and *brush* reference systems of PEG grafted to the anatase (1 0 1) TiO₂ surface, immersed in water and DCM. For water, also the number of *H-bonds* per monomer in the chain is reported and, for DCM, the number of $O_{PEG}$-Ti bonds per polymer.

| PEG@(1 0 1) | WATER | | | | DCM | | | |
|---|---|---|---|---|---|---|---|---|
| | MDFS (nm) | $\langle h^2 \rangle^{1/2}$ (nm) | $R_g$ (nm) | N° of H-bonds | MDFS (nm) | $\langle h^2 \rangle^{1/2}$ (nm) | $R_g$ (nm) | N° $O_{PEG}$-Ti bonds |
| mushroom | 1.04 ± 0.18 | 1.68 ± 0.49 | 0.67 ± 0.11 | 1.01 ± 0.35 | 0.77 ± 0.22 | 1.35 ± 0.55 | 0.58 ± 0.15 | 2.19 ± 0.68 |
| brush | 1.44 ± 0.01 | 2.00 ± 0.04 | 0.75 ± 0.01 | 0.52 ± 0.04 | 1.37 ± 0.01 | 1.93 ± 0.03 | 0.74 ± 0.01 | 1.04 ± 0.06 |

**Table 5**

Values of the *end-to-end distance* $\langle h^2 \rangle^{1/2}$ and *radius of gyration* $R_g$ (in nm) and number of *H-bonds* per monomer of the *mushroom*, *brush* reference model and PEG@NP systems, immersed in water at different grafting densities ($\sigma$).

| N. of chains Grafted | Grafting density $\sigma$ (chain/nm$^2$) | $\langle h^2 \rangle^{1/2}$ (nm) | $R_g$ (nm) | N. of H-bonds |
|---|---|---|---|---|
| 0 | Free | 1.43 ± 0.40 | 0.60 ± 0.08 | 0.99 ± 0.08 |
| mushroom | 0.107 | 1.68 ± 0.49 | 0.67 ± 0.11 | 1.01 ± 0.35 |
| 5 | 0.225 | 1.62 ± 0.16 | 0.64 ± 0.04 | 1.11 ± 0.07 |
| 10 | 0.440 | 1.68 ± 0.10 | 0.66 ± 0.02 | 1.03 ± 0.06 |
| 15 | 0.676 | 1.77 ± 0.12 | 0.67 ± 0.02 | 0.93 ± 0.05 |
| 20 | 0.901 | 1.73 ± 0.09 | 0.66 ± 0.02 | 0.92 ± 0.04 |
| 25 | 1.126 | 1.70 ± 0.06 | 0.65 ± 0.01 | 0.90 ± 0.04 |
| 30 | 1.351 | 1.81 ± 0.09 | 0.68 ± 0.02 | 0.85 ± 0.03 |
| 50 | 2.252 | 1.82 ± 0.05 | 0.68 ± 0.02 | 0.75 ± 0.03 |
| brush | 2.573 | 2.00 ± 0.04 | 0.75 ± 0.01 | 0.52 ± 0.04 |

**Table 6**

Values of the *mean distance from the surface* (MDFS) in nm of the *mushroom*, *brush* reference model and PEG@NP systems, immersed in DCM at different grafting densities ($\sigma$).

| N. of chains grafted | Grafting density $\sigma$ (chain/nm$^2$) | MDFS (nm) |
|---|---|---|
| MUSHROOM | 0.107 | 0.77 ± 0.22 |
| 5 | 0.225 | 0.43 ± 0.02 |
| 10 | 0.440 | 0.45 ± 0.01 |
| 15 | 0.676 | 0.58 ± 0.02 |
| 20 | 0.901 | 0.63 ± 0.02 |
| 25 | 1.126 | 0.70 ± 0.02 |
| 30 | 1.351 | 0.78 ± 0.03 |
| 50 | 2.252 | 0.96 ± 0.01 |
| brush | 2.573 | 1.37 ± 0.01 |

### 3.4. PEG grafted to curved TiO$_2$ nanoparticles:

#### 3.4.1. In water

In this section, we present the results for the PEG@NP system surrounded by a bulk of water. In Table 4, we report the *mean distance from the surface* (MDFS) of the PEG at different coverage densities. The values are around 1.00 nm for all the PEG@NP systems from $\sigma$ = 0.225 up to $\sigma$ = 1.351, indicating that there is no evident effect of the grafting density on this indicator and that the PEG chains are possibly in the same or in a similar conformation. When the density is very high ($\sigma$ = 2.252), the MDFS increases too, since the chains are now close one to the other and tend to stretch. For coverages up to $\sigma$ = 1.351 the MDFS values for the PEG@NP systems are similar to that for the *mushroom* model and they are all lower than for the *brush* one. It can be noticed that, despite the close grafting density value ($\sigma$ = 2.252 vs $\sigma$ = 2.573), the MDFS of the polymers grafted on the NP is much lower than the one of the *brush* polymers grafted on the flat surface. This is due to the fact that, while for flat surfaces the volume available per polymer chain is constant moving away from the surface, for the curved NP it increases, allowing the PEG to coil and arrange closer to the surface. This means that the PEGs around the NP reach a proper brush conformation at much higher densities with respect to the flat surface case, in line with what already observed for PEG grafted gold nanoparticles [62].

As for the case of the free PEG in water and on the flat (1 0 1) surface, we have evaluated the root mean square *end-to-end distance* $\langle h^2 \rangle^{1/2}$, the *radius of gyration* $R_g$ and the number of *H-bonds* per monomer for all the PEG@NP systems, as defined in Section 2.2. The values are reported in Table 5 and compared with those for the free polymer, the *mushroom* and the *brush* reference systems in water from previous sections. Distribution of the *radius of gyration* $R_g$ and of the number of *H-bonds* per monomer for the two grafting density extremes ($\sigma$ = 0.225 and $\sigma$ = 2.252) are reported in Fig. S5.

The *end-to-end distances* and *radii* of *gyration* for PEG@NP models are only slightly bigger with respect to the free PEG and they increase very little with the grafting density. This is a further confirmation that the grafted chains behaves almost as the free one in water, even at high coverage, at least till $\sigma$ < 1.351, above which a certain increase in the $\langle h^2 \rangle^{1/2}$ and $R_g$ can be noticed. Regarding the number of *H-bonds* established with the solvent molecules, at low coverage of PEG on the NP, it is higher than for the free chain and for the *mushroom* reference conformation on the flat surface. This is due to the long-range effect that a curved TiO$_2$ surface has on the surrounding water layers, which show a higher density with respect to the bulk value [57]. However, as the PEG coverage on the NP increases, the volume accessible to water in the polymer corona area around the nanoparticle decreases, reducing the number of established *H-bonds* per monomer. The reduction of H-bonds for the PEG@NP with the highest density ($\sigma$ = 2.252) is, however, much lower than for the *brush* reference model ($\sigma$ = 2.573) from the previous section. As discussed above, this means that, although the $\sigma$ values are similar, there is much more interaction between the polymers grafted to the flat surface and that the flat surface is much less accessible to water than the curved one.

To give a quantitative evaluation of the conformation of the free PEG and the grafted ones in water, we have calculated the *OCCO dihedral angle index* (DAI), as discussed in previous sections. In Fig. 4 the DAI is reported as a function of the grafting densities for the free PEG, for all the PEG@NP models, and for the *mushroom* and *brush* reference systems from the previous section. As we have already discussed in Section 3.3, the polymer in the *mushroom* reference model shows the same conformation of the free PEG (relatively small DAI value) indicating a strong PEG/SOLVENT interaction, while, on the other hand, in the *brush* case the DAI value is quite high, as one would expect for stretched chains that mainly interact among themselves. In the case of the PEG@NP systems, for low coverage ($\sigma$ < 0.440) the DAI remains very similar to

**Table 7**

Values of the *end-to-end distance* $\langle h^2 \rangle^{1/2}$ and *radius of gyration* $R_g$ (in nm) and number of $O_{PEG}$-Ti bonds per polymer of the PEG@NP systems, immersed in DCM at different grafting densities ($\sigma$).

| N. of chains grafted | Grafting density $\sigma$ (chain/nm$^2$) | $\langle h^2 \rangle^{1/2}$ (nm) | $R_g$ (nm) | N. $O_{PEG}$-Ti bonds |
|---|---|---|---|---|
| 0 | Free | 0.34 ± 0.03 | 0.51 ± 0.02 | – |
| Mushroom | 0.107 | 1.35 ± 0.55 | 0.58 ± 0.15 | 2.19 ± 0.68 |
| 5 | 0.225 | 1.22 ± 0.04 | 0.57 ± 0.01 | 2.60 ± 0.19 |
| 10 | 0.440 | 1.10 ± 0.02 | 0.57 ± 0.01 | 1.89 ± 0.15 |
| 15 | 0.676 | 1.32 ± 0.12 | 0.58 ± 0.01 | 1.62 ± 0.08 |
| 20 | 0.901 | 1.13 ± 0.04 | 0.59 ± 0.01 | 1.35 ± 0.09 |
| 25 | 1.126 | 1.21 ± 0.05 | 0.60 ± 0.01 | 1.14 ± 0.03 |
| 30 | 1.351 | 1.31 ± 0.06 | 0.64 ± 0.01 | 1.13 ± 0.03 |
| 50 | 2.252 | 1.51 ± 0.03 | 0.65 ± 0.01 | 1.03 ± 0.01 |
| Brush | 2.573 | 1.93 ± 0.03 | 0.74 ± 0.01 | 1.04 ± 0.06 |

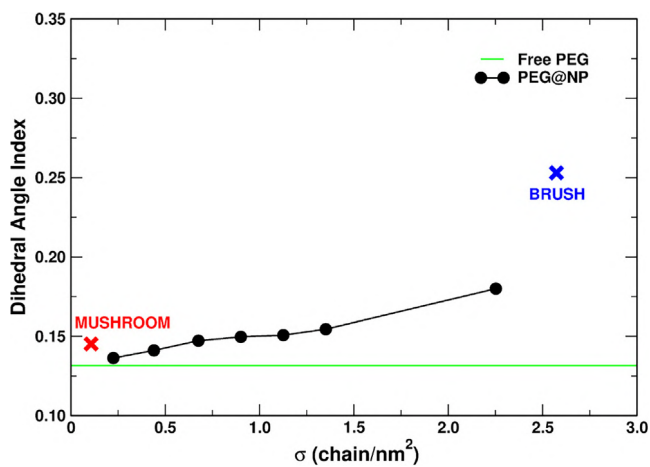

**Fig. 4.** *Dihedral Angle Index* (DAI) calculated for the free PEG chain (green line), the *mushroom* (red cross), *brush* (blue cross) reference systems and for all the different grafting densities of PEG@NP systems (black dots) in water. (For interpretation of the references to colour in this figure legend, the reader is referred to the web version of this article.)

that for the *mushroom* reference system, and thus to that for the free PEG. Surprisingly, also the PEG@NP systems with grafting densities up to $\sigma = 1.351$ show very similar conformation to the free PEG: the DAI slightly increases up to $\sigma = 1.351$, but it is still very far from the *brush* value. Only for the very high coverage ($\sigma = 2.252$), we observe an increased value of DAI and we can, therefore, deduce that a transition towards the *brush* conformation has occurred.

As a further step, we have assessed the applicability of the Daoud and Cotton model [97], for the conformation of star-like polymers, to our system. The model, which has been successively supported and extended by both analytical and experimental studies [98–102], predicts that, for high grafting density, spherical brushes present a central rigid core of polymer constant density, i.e. a concentrated regime where the volume fraction $\Phi(r)$ of the polymer changes as $\Phi(r) \propto r^{-1}$ and a semidilute regime where $\Phi(r) \propto r^{-4/3}$. In Fig. 5, we report the log-log plot of the PEG volume fraction as a function of the radial distance from the center of the NP for the different grafting densities ($\sigma$) of PEG@NP systems. As it can be observed, the $\Phi(r)$ behaviour for the highest grafting density ($\sigma = 2.252$) follows the Daoud-Cotton model, at least for distances up to $\sim$2.5 nm. This is not true for lower grafting densi-ties, as one would expect. Our results are in line with a recent study on PEGylated spherical gold nanoparticles, even if longer polymer chains were considered [62].

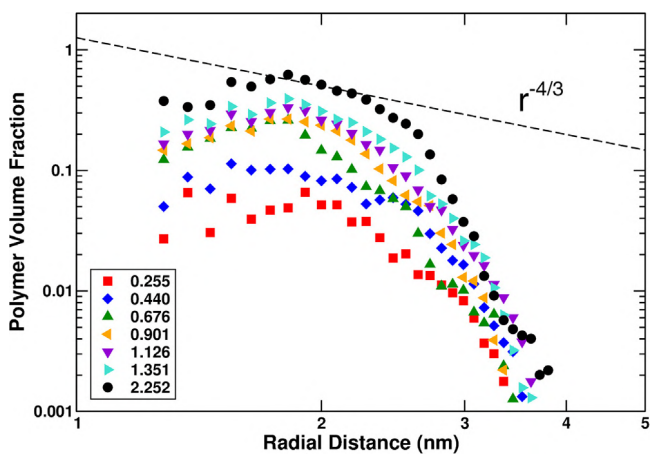

**Fig. 5.** Volume fraction $\Phi(r)$ of PEG around the nanoparticle in water as a function of the radial distance (in nm) from the NP baricenter. Different symbols indicate different grafting density values for the PEG@NP systems.

The last indicator analyzed in this work is the number of $O_{PEG}$-Ti bonds per PEG chain. For all the systems in this section, this value is 1, corresponding to the bond formed at the grafting site of each polymer. We have a further confirmation that both the strong NP/SOLVENT and PEG/SOLVENT interactions impede any further interaction between the surface and the backbone of the PEG chain.

Therefore, from what we have seen so far, we can safely infer that: (1) PEG strongly interacts with water establishing several H-bonds; (2) PEG does not interact with the NP surface apart from the grafting site and behaves as a free PEG chain up to very high densities (at least $\sigma = 1.351$); (3) water molecules bind the NP surface (as we have seen in Section 3.1) impeding the interaction between the backbone of the polymer and the NP; (4) due to the highly curved surface, only at extremely high coverage ($\sigma > 1.351$) the polymer chains start to interact more among themselves than with water, leading to the transition from the *mushroom* to the *brush* conformation; (5) in the *brush* model regime, the Daoud-Cotton model is respected.

### 3.4.2. In dichloromethane

In this section, we will investigate how the same systems of Section 3.4.1 behave in a different aprotic solvent: dichloromethane (DCM). In Table 6, we report the MDFS of PEG at different coverage densities on flat and curved surfaces. In DCM, the MDFS difference between the *mushroom* and the *brush* reference systems is slightly larger than in water. As already discussed in Section 3.3, the isolated polymer chain is more attracted by the flat (1 0 1) surface and remains quite close to it during the simulation run. On the other hand, at high coverage the PEG chains interact with themselves and remain, on average, farther from the flat (1 0 1) surface. Regarding the PEG@NP systems, at low-medium coverage ($\sigma < 1.126$) the polymer chains are very close to the surface, even more than for the *mushroom* reference system, indicating a larger interaction between the polymer and the curved surface of the NP with respect to a flat one. At very high density on the NP ($\sigma = 2.252$), the PEG chains arrange at larger distances from the surface, resulting in an MDFS value closer to the one for the *brush* reference model.

The *end-to-end distances* and *radii of gyration* of the PEG@NP in DCM are reported in Table 7. In this case, while the $R_g$ varies in a range similar to that observed for water, the $\langle h^2 \rangle^{1/2}$ behavior is quite different. The grafted polymers have much longer *end-to-end distances* than the free one in DCM, and this distance increases considerably with the coverage density indicating an overall stretching tendency of the PEG chains (especially for high $\sigma$ values). Distribution of the *radius of gyration* $R_g$ for the two grafting density extremes ($\sigma = 0.225$ and $\sigma = 2.252$) is reported in Fig. S5. In the case of DCM, we observe three different peaks indicating a competition between different polymer configurations. This is in contrast with the case of water, where the distribution of the *radius of gyration* (Fig. S5) is unimodal. It is very interesting to notice that in DCM there is a larger degree of interaction between the polymer and the NP's surface, as confirmed by the number of $O_{PEG}$-Ti bonds per PEG chain, which is larger than that observed for all the PEG@NP systems in water. This means that the oxygen atoms of the polymer chain can get so close to the NP surface to establish an electrostatic interaction with the undercoordinated Ti atoms. In particular, the lower the grafting density, the larger is the number of *bonds* (excluding the *mushroom* case where we have a less "reactive" flat (1 0 1) surface), in line with the shorter MDFS just mentioned above at the beginning of this section. As the grafting density increases, especially for $\sigma > 1$, the number of $O_{PEG}$-Ti bonds tends to 1 (this is the Ti—O interaction anchoring the PEG chain to titania), since the surface of the nanoparticle becomes inaccessible for multiple binding of a single chain, and the PEG@NP models start to resemble their analogous in water, i.e. a *brush-like* conformation.

We now compare the *OCCO dihedral angle index* (DAI) for all the investigated systems, as shown in Fig. 6. From Sections 3.2 and 3.3, we have learnt that the free PEG, the *brush* and the *mushroom* reference systems in DCM all present high DAI values. Therefore, the distribution of values is contained within a small range. In the case of the PEG@NP systems, the fact that DCM is a weakly interacting solvent, brings to the promotion of intramolecular (at very low coverage) or intermolecular (at higher coverage) PEG interactions. Also on the NP, at very low grafting density, linear segments within the polymer chains are formed (see Fig. S6 of the Supporting Material), resulting in high DAI values. Due to the aforementioned effects, all the PEG@NP systems have similar (close to *brush*) DAI values. Therefore, in the case of the DCM solvent, the DAI index is not clearly useful since it cannot discriminate between the different PEG conformation models.

The polymer volume fraction $\Phi(r)$, reported in Fig. 7, provides further evidence that at low coverage ($\sigma < 0.440$) the polymer is so close to the surface that $\Phi(r)$ goes to zero, already at ~2.5 nm from the NP center. As the grafting density increases, the $\Phi(r)$ decay behavior becomes progressively less steep. At the highest grafting density, $\Phi(r)$ presents an approximate $r^{-5/2}$ dependence up to 2.7 nm, which reflects the fact that the PEG chains stay quite far from the surface and corroborates what already concluded from the MDFS and from the number of $O_{PEG}$-Ti bonds. It is important to underline that in this case the Daoud-Cotton model is not followed and that $\Phi(r)$ still decreases very quickly compared to the corresponding high coverage systems in water. In other terms, the polymer corona height of the TiO₂ nanoparticle will be lower in DCM than in water, despite the same grafting density.

We may summarize the analysis in DCM solvent as follow: (1) the PEG/SOLVENT and the NP/SOLVENT interactions are very weak; (2) for this reason at low coverage (up to $\sigma = 0.440$) the polymer chains bind the undercoordinated Ti atoms of the NP surface with different oxygen atoms of their backbone (strong PEG/NP interaction); (3) at low coverage, the intramolecular PEG interaction induces the formation of linear polymer segments, which turns into higher DAI index values (numerous *trans* dihedral angles) than in water; (4) the MDFS increases considerably with the coverage, especially when $\sigma > 1$, where the interactions between the polymers became more relevant than the PEG/NP ones; (5) at high coverage the polymers adopt a *brush-like* conformation, which does not follow the Daoud-Cotton model.

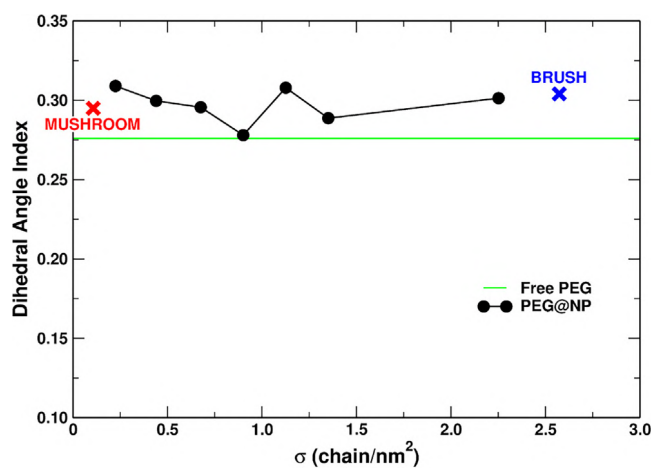

**Fig. 6.** *Dihedral Angle Index* (DAI) calculated for the free PEG chain (green line), for the *mushroom* (red cross), *brush* (blue cross) reference systems and for all the different grafting densities of PEG@NP systems (black dots) in DCM. (For interpretation of the references to colour in this figure legend, the reader is referred to the web version of this article.)

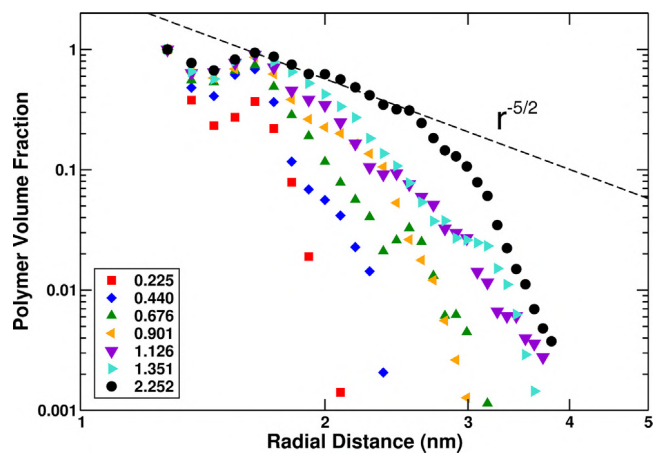

**Fig. 7.** Volume fraction $\Phi(r)$ of PEG around the nanoparticle in DCM as a function of the radial distance (in nm) from the NP baricenter. Different symbols indicate different grafting density values for the PEG@NP systems.

### 3.4.3. Comparison of solvent effects for water and DCM

In the previous two sections, we have analyzed the conformation of PEG@NP systems in different solvents, separately. Here, we will try to rationalize the differences in the observed polymers conformation in terms of the different nature of the two solvents considered: water and DCM.

From what we have seen so far, the PEG/SOLVENT interaction is stronger in the case of a protic solvent such as water, which can establish several H-bonds with the polymer chain (see Table 5 in Section 3.4.1), than for DCM, which behaves essentially as a $\Theta$ solvent. This is true not only when we consider the free polymer but also when PEG is grafted to the NP surface. Indeed, at low grafting density the PEG chains in water behave as the free one, while in DCM they bind to the NP surface mainly because they are not interacting with the solvent. In Fig. 8a we report the structure of low covered PEG@NP systems in water and DCM ($\sigma = 0.225$, corresponding to the dotted line in Fig. 8b), showing that, while in the first case the polymers are totally "immersed" in the solvent, in the case of DCM they are very close to the NP surface. As the coverage increases, the interaction between chains becomes more important than the one with the solvent, the polymers stretch and their MDFS increases, especially in the case of DCM where the MDFS rapidly tends to the one of PEG@NP in water (see Fig. 8b).

Considering the NP/SOLVENT interaction, we have seen (Fig. 1 in Section 3.1) that water arranges much closer than DCM around the curved surface of the nanoparticle since water molecules establish either coordinative bonds with the undercoordinated surface Ti atoms or H-bonds with surface O atoms and form few surrounding water layers that are denser than bulk water. For this reason, there is a strong competition between water and polymer adsorption on the surface, which reduces the possibility to create $O_{PEG}$-Ti bonds with the polymer backbone at any PEG@NP coverage. On the contrary, DCM is weakly interacting with the NP, therefore at low coverage the polymer can easily reach the surface without solvent shielding and establish lateral bindings. At medium-high coverage the situation is different since the PEG molecules start to interact among themselves and the formation of $O_{PEG}$-Ti bonds with the PEG backbone disappears, resulting in a number of $O_{PEG}$-Ti bonds very close to one (see Fig. 9a), i.e. only the anchoring Ti—O interaction.

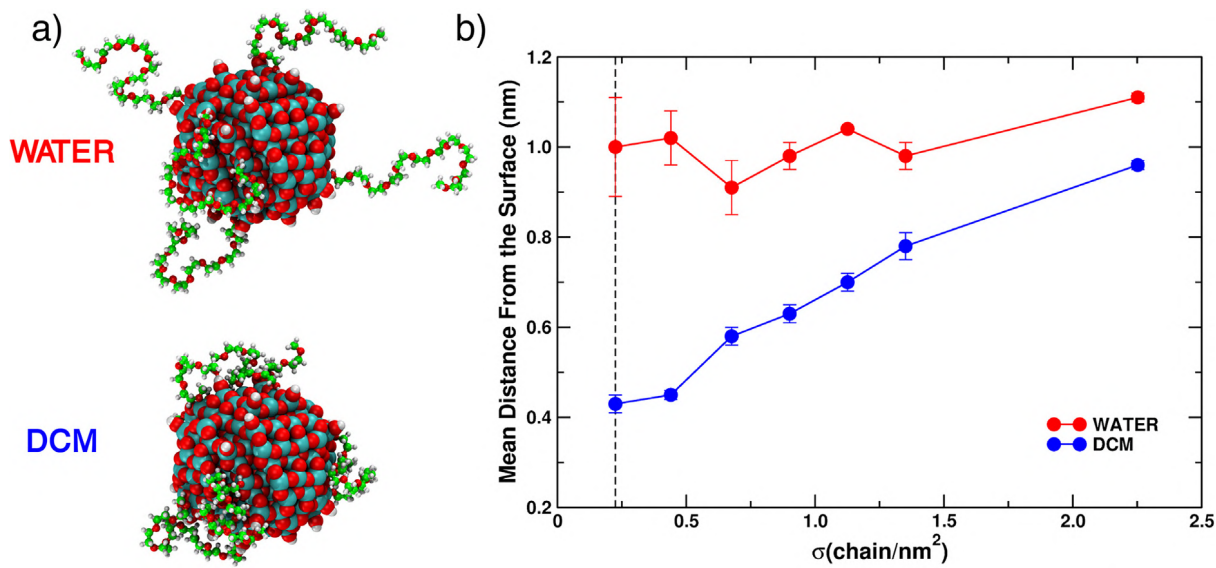

**Fig. 8.** (a) Last MD snapshots of the low-covered ($\sigma$ = 0.225) PEG@NP systems in water (top) and in DCM (bottom). Titanium, oxygen, carbon and hydrogen atoms are in cyan, red, green and white, respectively. (b) *Mean Distance From the Surface* (MDFS, in nm) as a function of the grafting density $\sigma$ (in chain/nm$^2$) for the PEG@NP systems in water (red) and DCM (blue). (For interpretation of the references to colour in this figure legend, the reader is referred to the web version of this article.)

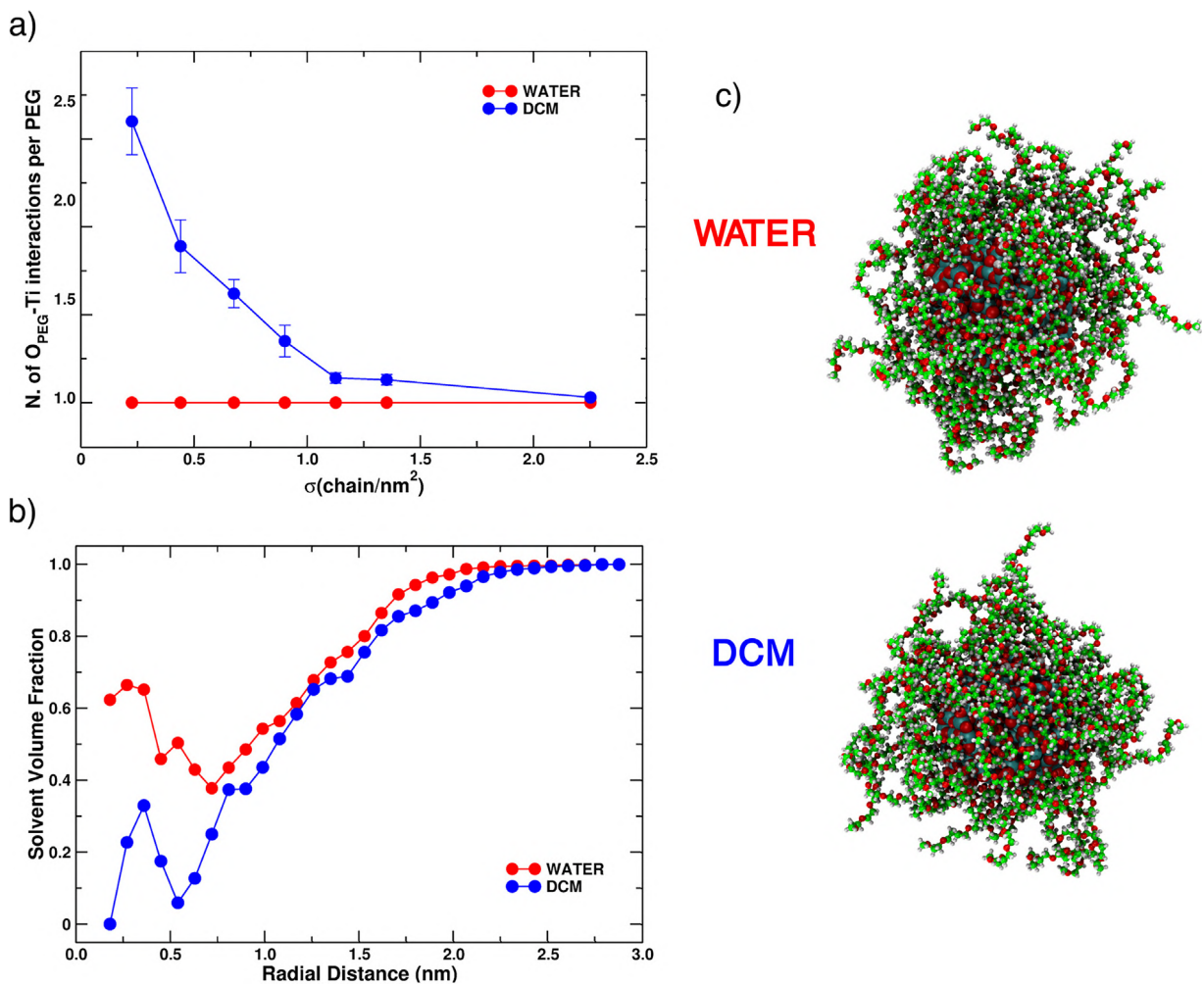

**Fig. 9.** (a) Number of $O_{PEG}$-Ti bonds per PEG as a function of the grafting density $\sigma$ (in chain/nm$^2$) for all the PEG@NP systems. (b) Volume fraction of the solvent around the PEG@NP with $\sigma$ = 2.252. Values in red and in blue refer to the systems in water and in DCM, respectively. (c) Last MD snapshots of the PEG@NP system at the highest grafting density ($\sigma$ = 2.252) in water (top) and in DCM (bottom). Titanium, oxygen, carbon and hydrogen atoms are in cyan, red, green and white, respectively. (For interpretation of the references to colour in this figure legend, the reader is referred to the web version of this article.)

In Fig. 9b, we report the water and DCM volume fraction for the higher coverage PEG@NP system considered ($\sigma = 2.252$, represented in Fig. 9c). Again, it is evident that water is more interacting with the NP surface. Indeed, even in the proximity of the surface (0 < r < 1 nm), there is a consistent volume of water (between 45 and 65%), while, in the same distance range, the volume of DCM is between 0 and 35%. This is also due to the much smaller volume per molecule for water (0.03 nm$^3$) with respect to DCM (0.06 nm$^3$), which allows the water molecules to better penetrate the polymer brush.

Finally, we have estimated the average polymer height ($H$) from the surface all around the NP from the *end-to-end distance* [62]. In Fig. 10, the log-log plot of the $\langle h^2 \rangle^{1/2}$ as a function of the grafting density $\sigma$ is reported both for water and DCM. Regarding water, the height follows the $\sigma^{1/5}$ scaling, which is what the Daoud-Cotton and other analytical models predict for small nanoparticles in a good solvent [93,95]:

$$H \propto (\sigma N^3 R^2)^{1/5}$$

where $\sigma$ is the grafting density of the polymer, $N$ is the number of monomers in the polymer chain, $R$ is the radius of the nanoparticle.

For flat (1 0 1) surfaces or large nanoparticles, the same models predict a steeper increase, $\sigma^{1/3}$ scaling, of the brush height with the grafting density [95]:

$$H \propto (\sigma N^3)^{1/3}$$

where $\sigma$ is again the grafting density of the polymer and $N$ the number of monomers in the polymer chain.

Differently, according to a coarse-grained model proposed by Lo Verso et al. [60], the brush height of a star polymer in a $\Theta$ solvent follows an intermediate behavior between that of a very small nanoparticle and of a flat surface, namely $\sigma^{1/4}$ scaling. As it can be noticed in Fig. 10, our results for DCM are in line with this prediction, excluding the low coverage regime, where the multiple binding of the PEG chains on the NP surface disturbs the correct scaling.

In both solvents, at high PEG grafting density with $\sigma = 2.252$, we expect the nanoparticles to be highly dispersed, as a consequence of the low TiO$_2$ surface accessibility and of a pronounced height of the polymer corona around the nanoparticle. However, when comparing the details of the various parameters (for example the end-to-end distance or the mean distance from the surface) in the two solvents, as it has been done in Figs. 8–10, we may conclude that miscibility in water is even more favorable than in DCM.

## 4. Conclusions

In summary, in this work we have presented a conformational study, based on force-fields molecular dynamics simulations, of PEG chains grafted on highly curved TiO$_2$ NPs (2–3 nm) at different grafting density and in different (protic and aprotic) solvents, i.e. water and DCM.

We compare the coating polymer conformations on spherical NPs with those on (1 0 1) flat surfaces, that are used as reference systems. Several indicators point to the fact that the grafted PEG chains strongly interact with water molecules in contrast with what happens in the DCM solvent.

In particular, we observe that in water the transition from *mushroom* to *brush* polymer conformation starts only at high density ($\sigma = 2.25$ chains/nm$^2$). In dichloromethane (DCM), at low-medium coverage ($\sigma < 1.35$ chains/nm$^2$), several interactions between the PEG chains backbone and undercoordinated Ti atoms are established, whereas at $\sigma = 2.25$ chains/nm$^2$ the conformation clearly becomes brush-like. Finally, we demonstrate that spherical brushes at high grafting density in water, but not in DCM, follow the Daoud-Cotton (DC) classical scaling model for the polymer volume fraction dependence with the distance from the center, which has been developed to describe star-shaped polymer systems.

Our study will help other researchers in the rational design of efficient synthetic protocols to obtain biocompatible nanosystems for medical applications, with great benefits for the in vivo performances.

## Acknowledgments

The authors are grateful to Lorenzo Ferraro, Gianluca Fazio, Massimo Tawfilas and Laura Bonati. The project has received funding from the European Research Council (ERC) under the European Union's HORIZON2020 research and innovation programme (ERC Grant Agreement No [647020]).

## Appendix A. Supplementary material

Supplementary data to this article can be found online at https://doi.org/10.1016/j.jcis.2019.07.106.

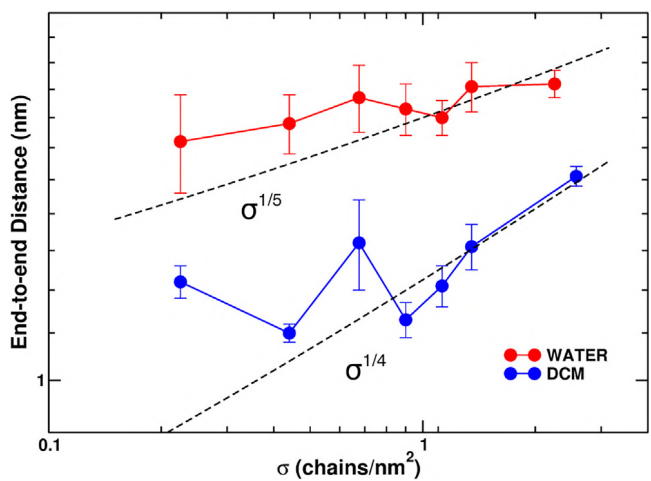

**Fig. 10.** *End-to-end distance* (in nm) as a function of the grafting density $\sigma$ (in chain/nm$^2$) for all the PEG@NP systems in water (red curve) and DCM (blue curve). (For interpretation of the references to colour in this figure legend, the reader is referred to the web version of this article.)

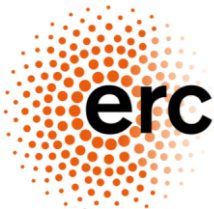

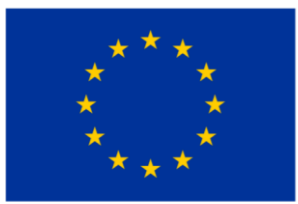

# Supporting Material

# Impact of Surface Curvature, Grafting Density and Solvent Type on the PEGylation of TiO$_2$ Nanoparticles

Daniele Selli[1], Stefano Motta[2], Cristiana Di Valentin[1]*

Dipartimento di Scienza dei Materiali, Università di Milano-Bicocca,
via R. Cozzi 55, I-20125 Milano, Italy
[2]Dipartimento di Scienze dell'Ambiente e della Terra, Università di Milano-Bicocca,
Piazza della Scienza 1, I-20126 Milano, Italy

* Corresponding author: cristiana.divalentin@unimib.it

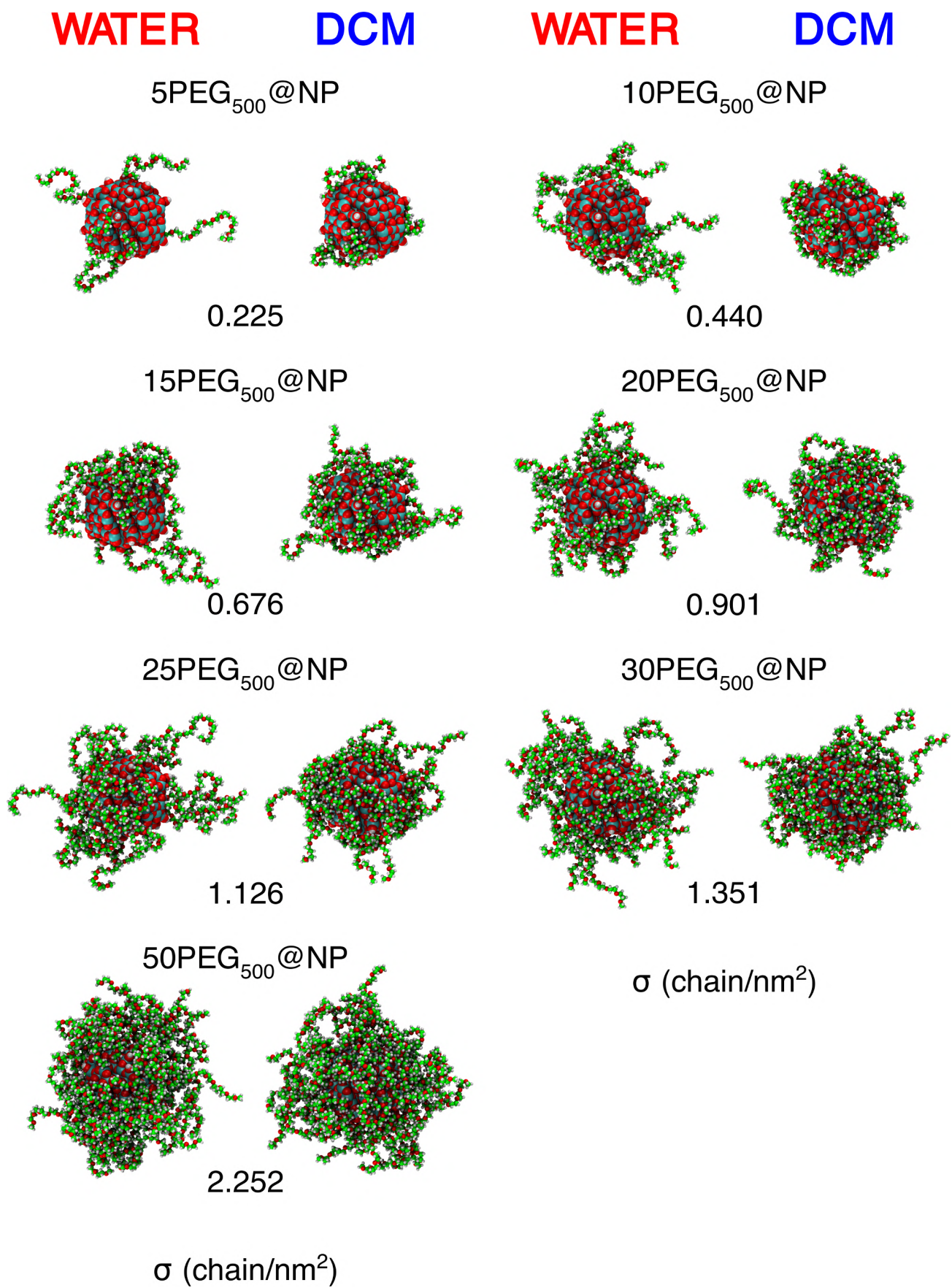

**FIGURE S1.** Final MD snapshot for the PEG@NP systems at different coverage ($\sigma$ in chains/nm$^2$) in water and DCM. The number of grafted PEG molecules to the NP is also reported. Titanium, oxygen, carbon and hydrogen atoms are in cyan, red, green and white, respectively. The water and DCM medium are not reported.

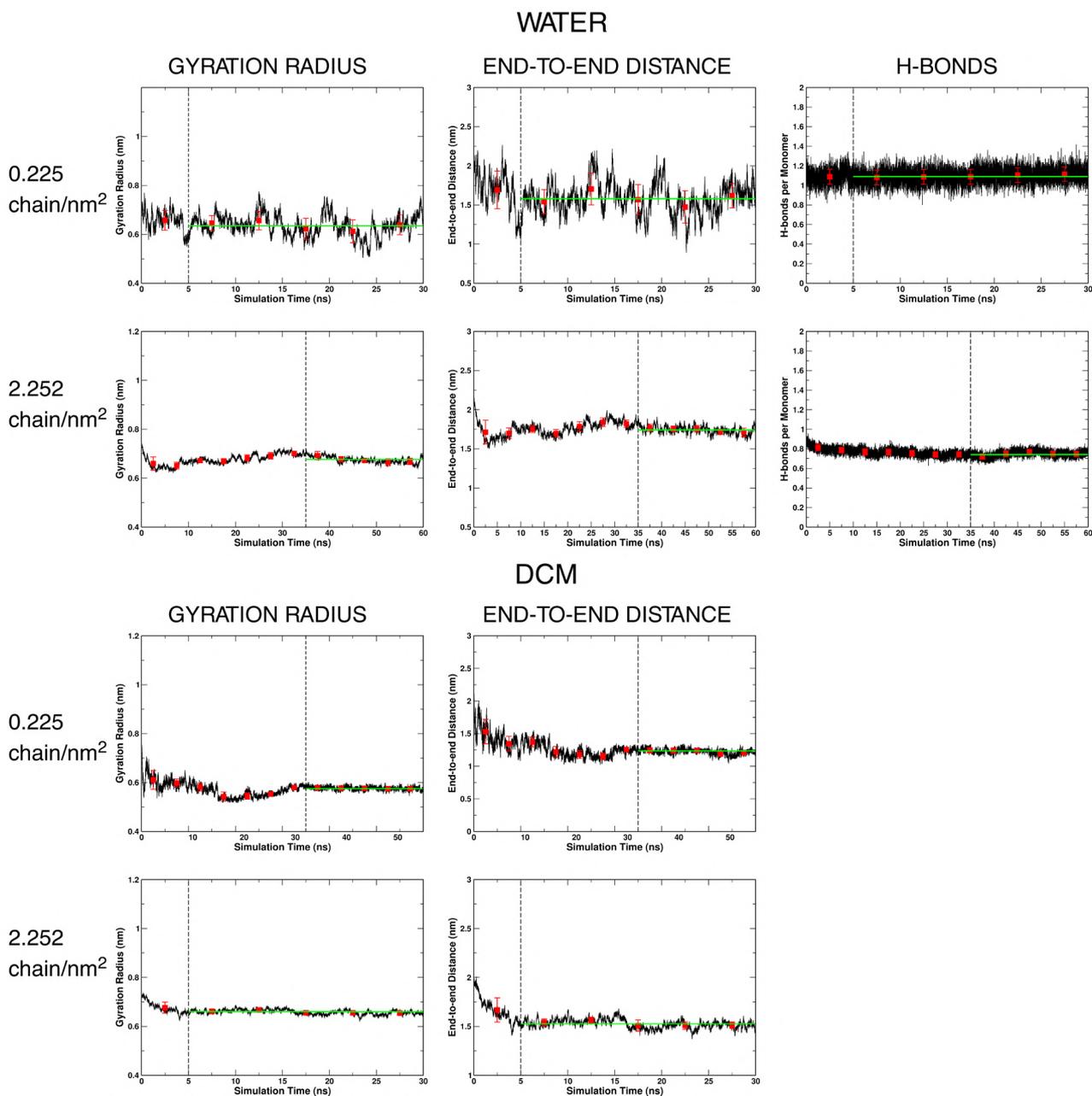

**FIGURE S2.** a) Radius of gyration, b) end-to-end distance and c) number of H-bonds per monomer for the PEG@NP ($\sigma$=0.225 and $\sigma$=2.252 chain/nm$^2$) system in water and DCM. The values of each indicator are collected every 5 ns during the equilibration (5 or 35 ns) and production (25 ns) runs. Red square are the values and relative standard deviation of each indicator averaged over 5 ns. The green line indicates the average of the indicators during the production run.

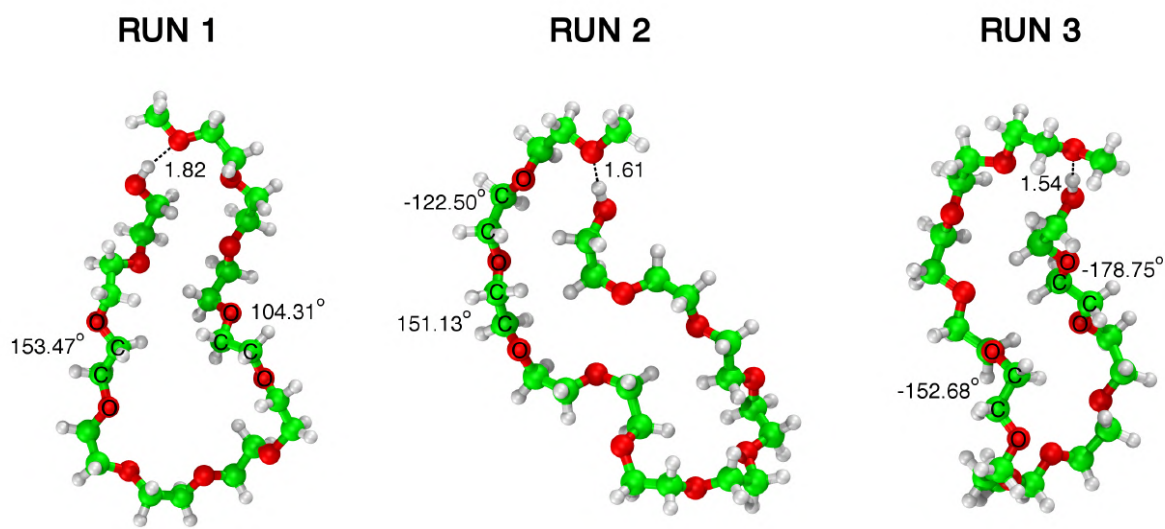

**FIGURE S3.** MD snapshots taken from the three different production run of the free PEG in DCM. For each snapshot intramolecular H-bonds are indicated with a dashed line (values in Å), while *OCCO dihedral angles* close to the *trans* value are indicated with the label of the four atoms (values in degree). High absolute values of the OCCO dihedral angles, result in linear segment of the polymer chain. Oxygen, carbon and hydrogen atoms are in red, green and white, respectively. The DCM medium is not reported.

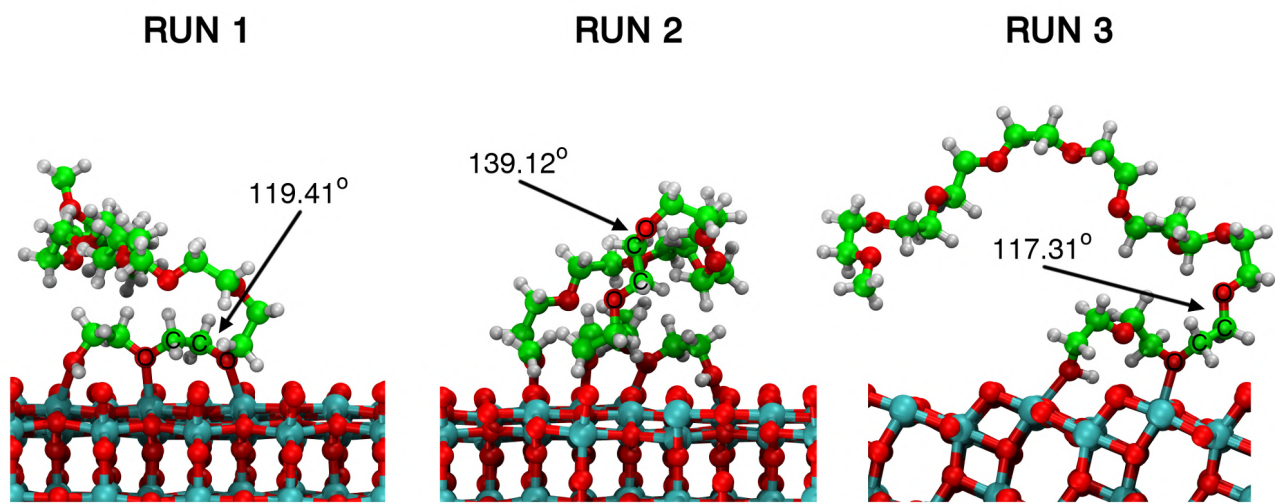

**FIGURE S4.** MD snapshots taken from the three different production run of the *mushroom* conformation of the PEG grafted on the anatase (101) $TiO_2$ surface in DCM. For each snapshot *OCCO dihedral angles* close to the *trans* value are indicated with the label of the four atoms (values in degree). High absolute values of the OCCO dihedral angles, result in linear segment of the polymer chain. Titanium, oxygen, carbon and hydrogen atoms are in cyan, red, green and white, respectively. The DCM medium is not reported.

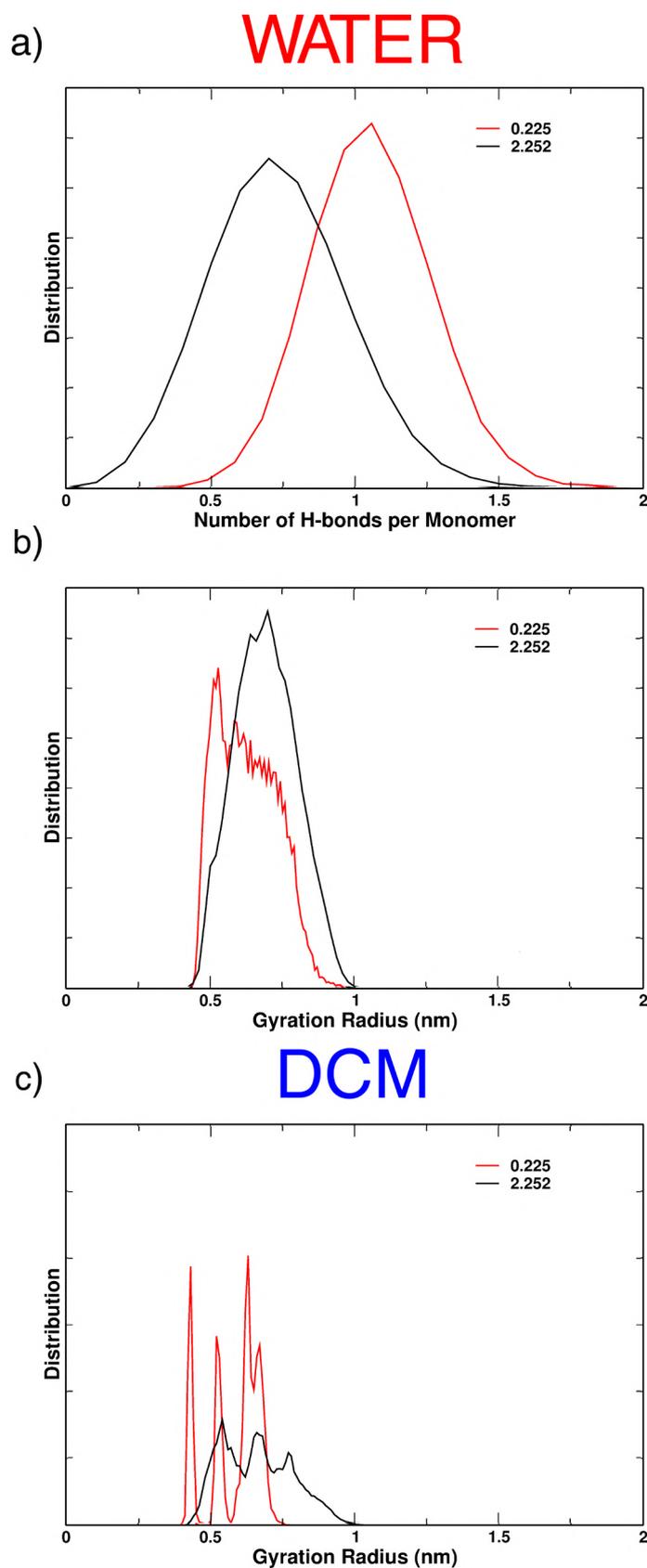

**FIGURE S5.** Distribution of a) number of H-bonds per monomer in water, b) radius of gyration in water and c) radius of gyration in DCM for PEG@NP systems with two grafting densities: σ=0.225 chain/nm² (red lines) and σ=2.252 chain/nm² (black lines).

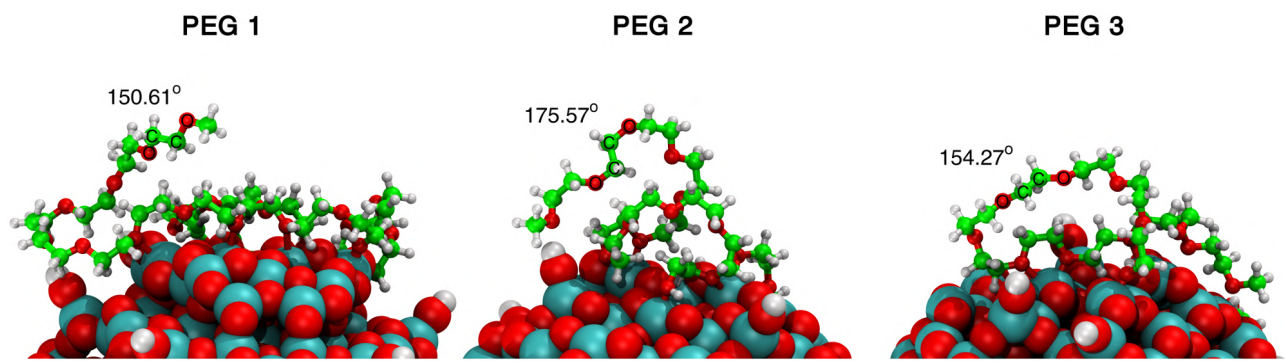

**FIGURE S6.** MD snapshots taken from the production run of the PEG@NP systems with grafting density σ=0.225 in DCM. For three different PEG (out of the five total ones), *OCCO dihedral angles* close to the *trans* value are indicated with the label of the four atoms (values in degree). High absolute values of the OCCO dihedral angles, result in linear segment of the polymer chain. Titanium, oxygen, carbon and hydrogen atoms are in cyan, red, green and white, respectively. The DCM medium is not reported.